\documentclass[12pt]{iopart}

\usepackage{iopams}
\usepackage{epsfig,multicol}
\begin{document}

\title[Colour Superconductivity in an external B]{Colour superconductivity in a strong magnetic field}

\author{Efrain J. Ferrer$^1$, Vivian de la Incera$^2$ and Cristina Manuel$^3$}

\address{$^{1,2}$Department of Physics, Western Illinois University, Macomb, IL 61455, USA}
\address{$^3$ Instituto de Fisica Corpuscular, CSIC-U. de Valencia, 46071 Valencia, Spain}
\ead{$^1$EJ-Ferrer@wiu.edu,$^2$V-Incera@wiu.edu,$^3$Cristina.Manuel@ific.uv.es}
\begin{abstract}
We explore
the effects of an applied strong external magnetic field in a three flavour
massless colour superconductor. The long-range component of the B field
that penetrates the superconductor enhances some quark condensates,
leading to a different condensation pattern. The external field also
reduces the flavour symmetries in the system, and thus it changes drastically
the  corresponding low energy physics. Our considerations are
relevant for the study of highly magnetized compact stars.

\end{abstract}


\section{Introduction}

This workshop is devoted to study the influence of extreme conditions on a  quantum field theory. In this talk
we will particularly focus our attention to the study of quantum chromodynamics (QCD) under the conditions of high baryonic density,
and
very strong magnetic fields.

After the discovery of the property of asymptotic freedom of QCD, it was soon realized that the behaviour of the theory
in the situations of very high temperature and/or baryonic density would be drastically different than in vacuum. In those circumstances
quarks and gluons should behave as almost free particles, as their coupling becomes very weak at high energy scales.
Currently the physics community is devoting a lot of efforts to test these ideas experimentally.

In different astrophysical settings, it is believed that the density is so high that the hadrons melt into
their fundamental constituents,
giving rise to quark matter. It has been known for long time now that cold dense quark matter should exhibit the phenomenon of
colour superconductivity \cite{Bailin-Love,reviews}. It is our aim here to explain how a strong magnetic
field affects this phenomenon. This is not
a simple academic question.
The real fact is that almost all compact stars sustain a strong magnetic field, of the order of $B \sim 10^{12} - 10^{14}$ G for pulsars,
and of $B \sim 10^{14}-10^{15}$ G for magnetars. A comparison of the gravitational and magnetic energies of a compact star tells us that the
maximum fields may be as high as $B \sim 10^{18}-10^{19}$ G.
The common believe is that all the above mentioned compacts objects are neutron
stars, where neutrons are in a superfluid phase, while the protons are in a superconducting one, probably
of type I.  In \cite{MCFL} it was found that an external magnetic field influences both quantitatively
and qualitatively the colour superconducting state. We will
 explain here why an applied external magnetic field has different effects in an electromagnetic or in a
colour superconductor. These considerations allow us to say that
if quark matter in a colour superconducting state is realized in any of those
compact objects, the dynamics of the associated magnetic field should be drastically different than in neutron matter. We hope
to explore the astrophysical consequences of our work in the near future.

\section{Electromagnetic and colour superconductivity}

Our present microscopic understanding of the phenomenon of colour superconductivity relies on techniques of
BCS theory adapted to dense quark
matter.

The seminal work of Bardeen, Cooper and Schrieffer gave the
microscopic explanation of the electromagnetic superconductivity
found in some metals at low temperature $T$. The main ingredients of
BCS theory are the following: a finite density of fermions at low
$T$, occupying a Fermi sea, and an attractive interaction between
them occurring close to the Fermi surface. Even if the interaction
is very weak, this will render the existing ground state unstable
favouring the formation of Cooper pairs of fermions. In a metal at
low temperature the attractive interaction is mediated through
phonons, the quantum of vibrational energy of the metal.

\begin{figure}[b]
\begin{center}
\mbox{ \psfig{file=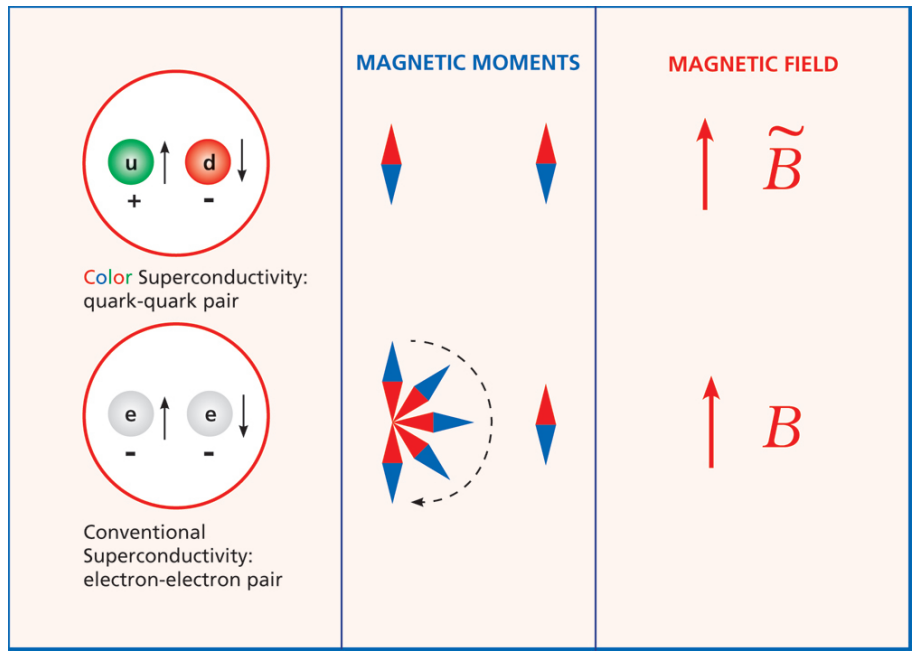,width=12cm} }
\caption{\label{mag-moments} The picture provides an intuitive way
to understand the magnetic reinforcement of colour
superconductivity.}
\end{center}
\end{figure}

In electromagnetic superconductivity, the Cooper pair, having two electrons (or any other two charged fermions)
has a net electric charge.
 The gauge symmetry is spontaneously broken, and the photon acquires a mass through
the Anderson-Higgs mechanism. A weak magnetic field is thus screened in the superconductor: this is the Meissner
effect. However, a sufficiently strong magnetic field destroys the superconducting state. One intuitive way to understand
why this is so is provided in the lower section of Fig.~\ref{mag-moments}. The electrons in the Cooper pair
have equal charges and opposite spins thus antiparallel magnetic moments. The field effect
will be to align the two magnetic moments parallel to each other,
therefore tending to break the superconducting bound state.
Electromagnetic superconductivity and
magnetism are thus anathema.

In high density QCD  quarks fill out the states up to the Fermi surface -the first
 required ingredient of BCS theory. The fundamental interaction
between two quarks mediated by one-gluon exchange
 has an attractive component in the colour antitriplet channel -the second required ingredient in BCS theory. At very large density the energy of the quarks at the Fermi surface
will be large, so the attractive interaction between them will be weak. According to BCS theory
even an arbitrarily
weak interaction will do the trick of restructuring the ground state through the formation of Cooper pairs
of quarks. Because the quarks carry "colour" charge, the quark-quark pairs will carry nonzero colour charge too,
and in general one may expect that some or all gluons may get a Meissner mass.

 Quarks carry different quantum
numbers (spin, colour, flavour). Who pairs with who? The answer depends on different QCD parameters, such
as the quark masses and chemical potentials. The last are further constrained by neutrality conditions.
 Then one talks about different colour superconducting phases,
which are characterized by the different local and global symmetries which are spontaneously broken.
In our analysis to study the influence of magnetism in the quark dynamics,
and as first approach to the problem, we will
neglect quark mass effects, and all subsequent complications, which we hope to address in future studies.

Quarks, apart from colour, also carry an electromagnetic charge. Should then one expect the colour
superconductor to be also an electromagnetic superconductor? The answer to this question requires
a closer look into the quark superconducting state. As we will explain below magnetism and colour superconductivity
are on good terms.

\section{Colour-flavour locking phase}

The ground state of QCD at high baryonic density with three
light quark flavours is described by the (spin zero) condensates \cite{alf-raj-wil-99/537}
\begin{equation}
  \langle q^{ai}_{L } q^{bj}_{L }\, \rangle
=-\langle q^{ai}_{R } q^{bj}_{R }\, \rangle
=   \Delta_A \, \epsilon^{abc} \epsilon_{ijc}
\ ,
\end{equation}
where $q_{L/R}$ are Weyl spinors (a sum over spinor indices is
understood), and $a,b$ and $i,j$ denote flavour and colour indices,
respectively. For simplicity we have neglected a colour sextet component of the condensate,
as it is a subleading effect.

The diquark condensates lock the colour and flavour transformations
breaking both- thus, the name colour-flavour locked (CFL) phase.
 The symmetry breaking pattern  in the CFL phase is
$
 SU(3)_C \times SU(3)_L \times SU(3)_R \times U(1)_B
\rightarrow SU(3)_{C+L+R}$.
There are only nine Goldstone bosons that survive to the
Anderson-Higgs mechanism. One is a singlet, scalar mode,
associated to the breaking of the baryonic symmetry, and the
remaining octet is associated to the axial $SU(3)_A$ group, just
like the octet of mesons in vacuum. At sufficiently high density, the anomaly is
suppressed, and then one can as well consider the spontaneous breaking of an approximated
$U(1)_A$ symmetry, and  an additional pseudo
Goldstone boson.

An important feature of spin-zero colour superconductivity is that
although the colour condensate has non-zero electric charge, there is
a linear combination of the photon $A_{\mu}$ and a gluon
$G^{8}_{\mu}$ that remains massless \cite{alf-raj-wil-99/537,
alf-berg-raj-NPB-02},

\begin{equation}
\widetilde{A}_{\mu}=\cos{\theta}\,A_{\mu}-\sin{\theta}\,G^{8}_{\mu} \ ,
\label{1}
\end{equation}
while the orthogonal combination
$\widetilde{G}_{\mu}^8=\sin{\theta}\, A_{\mu}+\cos{\theta}\,G^{8}_{\mu}$
is massive. In the CFL phase the mixing angle $\theta$ is
sufficiently small ($\sin{\theta}\sim e/g\sim1/40$). Thus, the
penetrating field in the colour superconductor is mostly formed by
the photon with only a small gluon admixture.

The unbroken $U(1)$ group corresponding to the long-range rotated
photon (i.e. the $\widetilde {U}(1)_{\rm e.m.}$) is generated, in
flavour-colour space, by $\widetilde {Q} = Q \times 1 - 1 \times Q$,
where $Q$ is the electromagnetic charge generator. We use the
conventions $Q = -\lambda_8/\sqrt{3}$, where $\lambda_8$ is the 8th
Gell-Mann matrix. Thus our flavour-space ordering is $(s,d,u)$. In
the 9-dimensional flavour-colour representation that we will use here
(the colour indexes we are using are (1,2,3)=(b,g,r)), the
$\widetilde{Q}$ charges of the different quarks, in units of
$\widetilde{e} = e \cos{\theta}$, are
\begin{equation}
\label{q-charges}
\begin{tabular}{|c|c|c|c|c|c|c|c|c|}
  \hline
  $s_{1}$ & $s_{2}$ & $s_{3}$ & $d_{1}$ & $d_{2}$ & $d_{3}$ & $u_{1}$ & $u_{2}$ & $u_{3}$ \\
  \hline
  0 & 0 & - & 0 & 0 & - & + & + & 0 \\
  \hline
\end{tabular}
\end{equation}

While a weak magnetic field only changes slightly the properties of the CFL superconductor,
in the presence of a strong magnetic field the condensation pattern is changed,
giving rise to a new phase, the magnetic colour-flavour locked (MCFL) phase.

\section{Magnetic colour-flavour locking phase}

An external magnetic field to the colour superconductor will be able to penetrate it in
the form of a ``rotated" magnetic field $\widetilde{B}$. With respect to this long-ranged field,
although all the superconducting pairs are neutral, a subset of them are formed by quarks with
opposite rotated $\widetilde{Q}$ charges.  Hence, it
is natural to expect that this kind of condensates will be
affected by the penetrating field, as the quarks couple minimally to the rotated
gauge field. Furthermore, one may expect that these condensates are
strengthened by the penetrating field, since their paired quarks,
having opposite $\widetilde{Q}$-charges and opposite spins, have
parallel (instead of antiparallel) magnetic moments (see the upper section of Fig.~\ref{mag-moments}).
 The situation
here has some resemblance to the magnetic catalysis of chiral
symmetry breaking \cite{MC}, in the sense that the magnetic field
strengthens the pair formation. Despite this similarity, the way the
field influences the pairing mechanism in the two cases is quite
different as we will discuss later on.

A strong magnetic field affects the flavour symmetries of QCD, as
different quark flavours have different electromagnetic charges.
For three light quark flavours, only the subgroup of $SU(3)_L
\times SU(3)_R$ that commutes with $Q$, the electromagnetic
generator, is a symmetry of the theory. Equally, in the CFL phase
a strong $\widetilde{B}$ field should affect the symmetries in the
theory, as $\widetilde{Q}$ does not commute with the whole locked
$SU(3)$ group. Based on this considerations, we proposed the
following diquark (spin zero) condensate \cite{MCFL}
\begin{equation}
  \langle q^{ai}_{L } q^{bj}_{L }\, \rangle
=-\langle q^{ai}_{R } q^{bj}_{R }\, \rangle
=   \Delta_A \, \epsilon^{ab3} \epsilon_{ij3} + \Delta_A^B \left( \epsilon^{ab2} \epsilon_{ij2} +
\epsilon^{ab1} \epsilon_{ij1} \right)
\ ,
\end{equation}
and as for the CFL case, we have only considered the leading antritiplet colour channel.
For a discussion of the remaining allowed structures in the subleading sextet channel see \cite{MCFL}.
Here we have been guided by the principle of highest symmetry, that is, the pair
condensation should retain the highest permitted degree of symmetry,
as then quarks of different colours and flavours will participate in
the condensation process to guarantee a maximal attractive channel
at the Fermi surface \cite{alf-raj-wil-99/537}.

In the MCFL phase
the symmetry breaking pattern is
$
SU(3)_C \times SU(2)_L \times SU(2)_R \times U(1)^{(1)}_A\times U(1)_B \times U(1)_{\rm e.m.}
\rightarrow SU(2)_{C+L+R} \times {\widetilde U(1)}_{\rm e.m.}
$.
Here the symmetry group $U(1)^{(1)}_A$ is related to a current
which is an anomaly free linear combination of $u,d$ and $s$ axial
currents, and such that  $U(1)^{(1)}_A \subset SU(3)_A$. The locked $SU(2)$ group
corresponds to the maximal unbroken symmetry, such that it maximizes the condensation
energy. The counting of broken generators, after taking into account the
Anderson-Higgs mechanism, tells us that there are only five Goldstone
bosons. As in the CFL case, one is associated to the breaking of
the baryon symmetry; three Goldstone bosons are associated to the
breaking of $SU(2)_A$, and another one associated to the breaking
of  $U(1)^{(1)}_A$. If the effects of the anomaly could
be neglected, there would be another pseudo Goldstone
boson associated to the $U(1)_A$ symmetry.

To study the MCFL phase we used a Nambu-Jona-Lasinio (NJL)
four-fermion interaction abstracted from one-gluon exchange
\cite{alf-raj-wil-99/537}. This simplified treatment, although
disregards the effect of the $\widetilde {B}$-field on the gluon
dynamics and assumes the same NJL couplings for the system with
and without magnetic field, keeps the main attributes of the
theory, providing the correct qualitative physics.  The NJL
model is treated as the proper effective field theory to study colour
superconductivity in the regime of moderate densities.
The model is defined by two parameters, a coupling constant
$g$ and an ultraviolet cutoff $\Lambda$. The cutoff should be much
higher than the typical energy scales in the system, that is, the
chemical potential $\mu$ and the magnetic energy
$\sqrt{\widetilde{e}\widetilde{B}}$.

The study of the MCFL gap equations presents several technical
difficulties, which we will only briefly mention here. The
computation of the field-dependent quark propagators is laborious,
but it can be managed with the use of the Ritus' method,
originally developed for charged fermions \cite{Ritus:1978cj} and
recently extended to charged vector fields \cite{efi-ext}. In
Ritus' approach the diagonalisation in momentum space of charged
fermion Green's functions in the presence of a background magnetic
field is carried out using the eigenfunction matrices $E_p(x)$.
These are the wave functions of the asymptotic states of charged
fermions in a uniform magnetic field and play the role in the
magnetized medium of the usual plane-wave (Fourier) functions
$e^{i px}$ at zero field. With the help of the $E_p(x)$ functions,
one can compute the propagators in momentum space, which depend on
a discrete index that labels the Landau levels. With these
propagators, one can derive the MCFL gap equations.

The gap equations for arbitrarily value of the magnetic field
are extremely difficult to solve, and they require a numerical treatment.
However, we have found a situation where an analytical solution can
be found. This corresponds to the case
 $\widetilde{e}\widetilde{B} >\mu^2/2$, where $\mu$ is the chemical potential. In this case, only
charged quarks in the lowest Landau level contribute to the gap equation,
a situation that drastically simplifies the analysis.

In BCS theory, and in the presence of contact interactions, the fermionic gap has an
exponential dependence on the inverse of the density of states close to the
Fermi surface, which is proportional to $\mu^2$.
Effectively, one can find that within the NJL model the CFL gap reads

\begin{equation}
\label{gapCFL}
\Delta^{\rm CFL}_A \sim 2 \sqrt{\delta \mu} \, \exp{\Big( -\frac{3
\Lambda^2 \pi^2} {2 g^2 \mu^2} \Big) } \ .
\end{equation}
with $\delta \equiv \Lambda - \mu$.
In the MCFL phase, when $\widetilde{e}\widetilde{B} >\mu^2/2$, we find instead
\begin{equation}
\label{gapBA}
\Delta^B_A \sim 2 \sqrt{\delta \mu} \, \exp{\Big( - \frac{3 \Lambda^2
\pi^2} {g^2 \left(\mu^2 + \widetilde{e} \widetilde{B} \right)}
\Big) } \ .
\end{equation}
For the value of the remaining gaps of the MCFL phase, see
\cite{MCFL}. All the gaps feel the presence of the external magnetic
field. As expected, the effect of the magnetic field in
$\Delta^{B}_{A}$ is to increase the density of states, which enters
in the argument of the exponential as typical of a BCS solution. The
density of states appearing in (\ref{gapBA}) is just the sum of
those of neutral and charged particles participating in the given
gap equation (for each Landau level, the density of states around
the Fermi surface for a charged quark is
$\widetilde{e}\widetilde{B}/2 \pi^2$). The gap formed by
$\widetilde{Q}$-neutral particles, although  modified by the
$\widetilde{B}$ field \cite{MCFL}, has a subleading effect in the
MCFL phase.

As mentioned at the beginning of this Section, the situation here
shares some similarities with the magnetic catalysis of chiral
symmetry breaking \cite{MC}; however, the way the field influences
the pairing mechanism in the two cases is quite different. The
particles participating in the chiral condensate are near the
surface of the Dirac sea. The effect of a magnetic field there is
to effectively reduce the dimension of the particles at the lowest
Landau level, which in turn strengthens their effective coupling,
catalyzing the chiral condensate. Colour superconductivity, on the
other hand, involves quarks near the Fermi surface, with a pairing
dynamics that is already $(1+1)$-dimensional. Therefore, the
${\widetilde B}$ field does not yield further dimensional
reduction of the pairing dynamics near the Fermi surface and hence
the lowest Landau level does not have a special significance here.
Nevertheless, the field increases the density of states of the
${\widetilde Q}$-charged quarks, and it is through this effect, as
shown in  (\ref{gapBA}), that the pairing of the charged particles
is reinforced by the penetrating magnetic field.

\section{Conclusions}

We have presented the arguments to explain why three light flavour colour superconductivity
is made stronger, not weaker, by the presence of magnetism. These arguments have been
corroborated by an explicit computation of the quark gaps within a NJL model, in the regime
of strong magnetic fields. To better understand the
relevance of this new phase in astrophysics we need to explore the
region of moderately strong magnetic fields
$\widetilde{e}\widetilde{B}< \mu^2/2$, which requires to carry out
a numerical study of the gap equations including the effect of
higher Landau levels.

The presence of a strong magnetic field affects the values of the quark
gaps, and thus, it will modify the equation of state of the colour superconductor,
although we do not expect this to be a very pronounced effect. More drastically,
the low energy physics of the MCFL phase would differ from that of the CFL phase,
through the disappearance of light degrees of freedom. This fact will have consequences
on several macroscopic properties of the superconductor, that we hope to explore
soon.

\ack

The
work of E.J.F. and V.I. was supported in part by NSF grant
PHY-0070986, and C.M. was supported by MEC under grant
FPA2004-00996.

\noappendix

\noappendix

\end{document}